\documentstyle[prd,aps,graphicx,twocolumn]{revtex}

\begin{document}

\draft



\def\al{\alpha}  
\def\be{\beta} 
\def\ga{\gamma}
\def\de{\delta}
\def\ep{\epsilon}
\def\ze{\zeta}
\def\et{\eta}
\def\th{\theta}
\def\io{\iota}
\def\ka{\kappa}
\def\la{\lambda}
\def\rh{\rho}
\def\si{\sigma}
\def\ta{\tau}
\def\up{\upsilon}
\def\ph{\phi}
\def\ch{\chi}
\def\ps{\psi}
\def\om{\omega}
\def\De{\Delta}
\def\Ga{\Gamma}
\def\Th{\Theta}
\def\La{\Lambda}
\def\Si{\Sigma}
\def\Up{\Upsilon}
\def\Ph{\Phi}
\def\Ch{\Chi}
\def\Ps{\Psi}
\def\Om{\Omega}

\def\varep{\varepsilon}

\def\be{{\mathbf{e}}}
\def\bg{{\mathbf{g}}}
\def\bh{{\mathbf{h}}}
\def\bj{{\mathbf{j}}}
\def\bk{{\mathbf{k}}}
\def\bn{{\mathbf{n}}}
\def\bp{{\mathbf{p}}}
\def\bq{{\mathbf{q}}}
\def\br{{\mathbf{r}}}
\def\bu{{\mathbf{u}}}
\def\bv{{\mathbf{v}}}
\def\bx{{\mathbf{x}}}
\def\by{{\mathbf{y}}}
\def\bz{{\mathbf{z}}}
\def\bA{{\mathbf{A}}}
\def\bB{{\mathbf{B}}}
\def\bJ{{\mathbf{J}}}
\def\bE{{\mathbf{E}}}
\def\bF{{\mathbf{F}}}
\def\bR{{\mathbf{R}}}
\def\bal{\mbox{\boldmath $\al$}}
\def\bbe{\mbox{\boldmath $\be$}}
\def\bsi{\mbox{\boldmath $\si$}}
\def\bna{\mbox{\boldmath $\na$}}
\def\bdot{\mbox{\boldmath $\cdot$}}
\newcommand{\bsp}[2]{\mbox{\boldmath $#1\cdot#2$}}

\def\N{{\cal N}}
\def\L{{\mathcal{L}}}
\def\H{{\mathcal{H}}}
\def\P{{\mathcal{P}}}


\newcommand{\ben}{\begin{equation}}
\newcommand{\een}{\end{equation}}
\newcommand{\bea}{\begin{eqnarray}}
\newcommand{\eea}{\end{eqnarray}}
\newcommand{\ba}{\begin{array}}
\newcommand{\ea}{\end{array}}
\newcommand{\bi}{\begin{itemize}}
\newcommand{\ei}{\end{itemize}}


\def\math{\mathsurround 0pt}
\def\oversim#1#2{\lower.5pt\vbox{\baselineskip0pt \lineskip-.5pt
        \ialign{$\math#1\hfil##\hfil$\crcr#2\crcr{\scriptstyle\sim}\crcr}}}
\def\lap{\mathrel{\mathpalette\oversim {\scriptstyle <}}}
\def\gap{\mathrel{\mathpalette\oversim {\scriptstyle >}}} 

\def\dbar{{{\,\mathchar'26}\mkern-9mu\!d\,}}
\def\debar{{{\,\mathchar'26}\mkern-9mu\delta}}

\def\no#1{\mbox{\boldmath $:$} #1 \mbox{\boldmath $:$}}

\def\pa{\partial}

\def\half{\frac{1}{2}}
\def\na{\nabla}
\def\ap{\approx}
\def\vp{\varphi}
\def\pt{\propto}
\def\tr{\mathop{\rm tr}\nolimits}
\newcommand{\ket}[1]{|#1\rangle}
\newcommand{\bra}[1]{\langle#1|}
\newcommand{\braket}[2]{\langle#1|#2\rangle}
\newcommand{\sla}[1]{\not\!#1}

\def\omk{{\omega_{\bk}}}
\def\omkp{{\omega_{\bk'}}}
\def\omp{\omega_{\mathbf{p}}}
\def\id{{\mathsf 1}}

\def\prll{\parallel}
\newcommand{\norm}[1]{\frac{1}{\sqrt{2\om_{#1}}}}

\twocolumn[\hsize\textwidth\columnwidth\hsize\csname@twocolumnfalse\endcsname

\title{Inverse cascade in decaying 3D magnetohydrodynamic turbulence} 

\author{Mattias Christensson\cite{mcaddress}, Mark Hindmarsh\cite{mhaddress}}
\address{Centre for Theoretical Physics, University of Sussex, Brighton BN1 9QJ, U.K.}
\author{Axel Brandenburg\cite{abaddress}}
\address{Nordita, Blegdamsvej 17, DK-2100 Copenhagen, Denmark; \\
Department of Mathematics, University of Newcastle, Newcastle upon Tyne, NE1 7RU, 
U.K.}

\maketitle

\begin{abstract} 
 We perform direct numerical simulations 
 of three-dimensional freely decaying magnetohydrodynamic (MHD) turbulence. 
 For helical magnetic fields an inverse cascade effect is observed in 
 which power is transfered from smaller scales to larger scales. 
 The magnetic field reaches a scaling regime with self-similar evolution,
 and power law behavior at high wavenumbers. We also find power law decay in the 
 magnetic and kinematic energies, and power law growth in the characteristic 
 length scale of the magnetic field.
\end{abstract}
\pacs{PACS numbers: 95.30.Q, 47.65, 47.40, 98.80
\hfill SUSX-TH/00-018
\hfill NORDITA-2000-103 AP
}
\vskip.5pc
]

\section{Introduction}
Within cosmology, astrophysics or geophysics one often needs to deal with 
electrically conducting plasmas
at high kinematic and magnetic Reynolds numbers where magnetic fields are
dynamically important. Indeed, much of the turbulence in the interstellar
medium is magnetohydrodynamic in nature. 

Hydromagnetic turbulence has been explored extensively in connection with 
the generation of large scale magnetic fields in astrophysical bodies
such as planets, stars, accretion discs and galaxies through dynamo theories.
Non-driven, freely decaying turbulence may also be of interest in connection
with both the physics of the interstellar medium and cosmology. Our interest 
was inspired by the cosmology of primordial magnetic fields, which
is sometimes considered as a possible source for providing the
seed for the galactic dynamo \cite{Zel88}.  

There have been various related works on decaying MHD turbulence, 
by authors interested in different contexts 
\cite{Hos+95,PolPouSul95,GalPolPou97,Mac+98,BisMul99,MulBis00}. 
Most directly comparable to our work, 
Biskamp and M\"uller \cite{BisMul99} studied the energy decay in incompressible
3D magnetohydrodynamic turbulence in numerical simulations at relatively high 
Reynolds number, and in a companion letter \cite{MulBis00} studied the 
scaling properties of the energy power spectrum.

We are here especially interested in the inverse cascade of magnetic helicity
whereby magnetic energy is transferred from small to large scale 
fluctuations. This is important for a primordial magnetic field to reach a 
large enough scale with sufficient amplitude to be relevant for seeding the 
galactic dynamo \cite{HinEve98}.

It should be noted that due to the conformal invariance of MHD in the 
radiation era the MHD equations in an expanding universe can be converted
into the relativistic MHD equations in flat spacetime by an appropriate scaling
of the variables and by using conformal time \cite{BraEnqOle96}.
The equations of \cite{BraEnqOle96} differ slightly from the ordinary 
non-relativistic MHD equations. However, in order to facilitate comparison 
with earlier work, we use the non-relativistic equations.

We perform 3D simulations both with and without magnetic helicity, 
starting from statistically homogeneous and isotropic random initial 
conditions, with power spectra suggested by cosmological applications. 
We find a strong inverse cascade in the helical case, with equivocal 
evidence for a weak inverse cascade when only helicity fluctuations are present. 
In the helical case we also find a
self-similar power spectrum with an approximately $k^{-2.5}$ behavior at 
high $k$.
We present energy decay laws which are 
comparable to those found in the incompressible case by Biskamp and 
M\"uller \cite{BisMul99}, and in the compressible case by 
Mac Low et al.\ \cite{Mac+98}.

\section{The model} 
 We consider the equations for an isothermal compressible gas with a 
 magnetic field, which is governed by the momentum equation, the continuity
 equation, and the induction equation, written here in the form
 \bea
    \frac{\pa\bu}{\pa t} = 
    -\bu\cdot\bna\bu - 
    c_{s}^{2}\bna \ln\rho +
    \frac{\bJ\times\bB}{\rho}\nonumber\\
    +\frac{\mu}{\rho}\left(\na^{2}\bu + 
    \frac{1}{3}\bna\bna\cdot\bu\right),
 \eea
 
 \ben
    \frac{\pa \ln\rho}{\pa t} = 
    -\bu\cdot\bna \ln\rho - 
    \bna\cdot\bu,
 \een
 
 \ben
    \frac{\pa\bA}{\pa t} = 
    \bu\times\bB + 
    \eta\na^{2}\bA,
 \een
 where $\bB=\nabla\times \bA$ is the magnetic field in terms of the magnetic
 vector potential $\bA$, $\bu$ is the velocity, $\bJ$ is the current density, 
 $\rho$ is the density, $\mu$ is the dynamical viscosity, and $\eta$ is the 
 magnetic diffusivity.
 
 The code for solving these equations \cite{Bra00}
 uses a variable third order Runge-Kutta
 timestep and sixth order explicit centered derivatives in space. 
 All our runs are performed on a $120^3$ grid, and 
 we use periodic boundary conditions,
 which means that the average plasma density
 $\langle \rho_{0} \rangle = \rho_{0}$ is conserved during runs.
 Here $\rho_{0}$ is the value of the initially uniform density, and the 
 brackets denote volume average. 
 
 We adopt nondimensional quantities by measuring $\bu$ in units of $c$, 
 where $c$ is the speed of light, $\bk$ in units of $k_{1}$, where $k_{1}$
 is the smallest wavenumber in the box, which has a size of 
 $L_{\rm BOX} = 2\pi$, density in units of $\rho_0=1$, and $\bB$ is measured
 in units of $\sqrt{\mu_{0}\rho_{0}} c$, where $\mu_{0}$ is the magnetic
 permeability. This is equivalent to putting 
 $c = k_{1} = \rho_{0} = \mu_{0} = 1$.
 In the following we will refer to the mean kinematic viscosity $\nu$ which
 we define as $\nu \equiv \mu/\rho_{0}$. The sound speed $c_{s}$ takes
 the value $c_s = 1/\sqrt{3}$, as appropriate for a relativistic fluid.
 With $c =1$, the unit of time is such that the light crossing time of the
 box is $2 \pi$. 
 
 Our equations are similar to those for the
 relativistic gas in the early universe using scaled variables and
 conformal time for non-relativistic bulk velocities \cite{BraEnqOle96}.
 We expect our results to change little using the true relativistic equations, 
 as our advection velocity is at most only mildly relativistic, and this only 
 at the beginning of the simulation.

\section{On the role of the inverse cascade} 
The magnetic helicity $H_{\rm M}$ is given by
 \ben
 H_{\rm M} = \int\bA\cdot\bB\, d^{3}x
 \een
 and characterizes the linkage between magnetic field lines. 
 $H_{\rm M}$ is conserved in the absence of ohmic dissipation,
 although  it is still possible to 
 have local, small scale helicity fluctuations.  
 Helicity plays an important role in dynamo theory 
\cite{PouFriLeo76,MenFriPou81},
where turbulence is driven.  
  
 In many astrophysical and cosmological situations the magnetic Reynolds 
 number $Re_{\rm M}$ is very large. We define
 the magnetic Reynolds number as $Re_{\rm M} = Lv/\eta$,
 where $L$ and $v$ are the typical length scale and velocity of the
 system under consideration and $\eta$ is the resistivity.
 The magnetic Reynolds number is a measure of the relative
 importance of flux freezing versus resistive diffusion. 
 In a cosmological context this number can be extraordinarily 
 large: causality imposes the weak limit $L \le ct$ and relativity 
 demands $v < c$. 
 With conductivities relevant to the era when the electroweak phase transition 
 took place \cite{AhoEnq96}, one can in principle 
 obtain a magnetic Reynolds number of about $10^{16}$.
 This is often taken to mean that the  magnetic field is frozen into 
 the plasma, and the scale length of the field increases only with the 
 expansion of the Universe.
 
 However, this simple picture does not necessarily give a full description
 of the dynamics because 
 the MHD equations, especially at high Reynolds numbers where nonlinear terms
 are important, exhibit turbulent behavior, which 
 can lead to a redistribution of magnetic energy over different
 length scales \cite{BraEnqOle96}.
 Energy in a turbulent magnetic field can undergo an inverse cascade 
 and be transferred from high frequency modes to low frequency modes, 
 increasing the overall comoving correlation length \cite{PouFriLeo76}.
 This process is due to the 
 nonlinear terms giving rise to interactions between many different length 
 scales.
 
 We will take the initial primordial power spectrum as given
 and address the question of how such a primordial 
 spectrum evolves as a consequence of the nonlinear equations of motion.

 \section{Initial conditions}
 Since one of the aims of the present work is to investigate the role of magnetic
 helicity in the inverse cascade we describe how the initial conditions 
 for our simulations were set up. We chose our initial condition by setting up 
 magnetic fluctuations with an initial power spectrum 
 $P_{\rm M}(k)\equiv \langle\bB^{*}_\bk\cdot\bB_\bk\rangle\ap k^{n}$ in Fourier 
 space (and averaged over shells of constant $k=|\bk|$),
 for low values of the wavenumber $k$, using a exponential cutoff $k_c$. 
[The shell-averaged power spectrum, $P_{\rm M}(k)$, is not to be confused with the
shell-integrated energy spectrum, $E_{\rm M}=4\pi k^2\times{1\over2}P_{\rm M}(k)$,
which is shown in the plots below.]

 The magnetic field fluctuations are drawn from a Gaussian random field
 distribution fully determined by its power spectrum in Fourier space according to 
 the following procedure. For each grid point we use the corresponding wavenumber
 to select an amplitude from a Gaussian distribution centered on zero and with the
 width 
\ben
P_{\rm M}(k)=P_{\rm M,0}k^{n} \exp ( - (k/k_c)^4)
\een
where $k=|\bf k|$. 
 We then transform the field back into real space to obtain the 
 field at each grid point. This is done independently for each field component.

 There is a requirement in cosmology that $n \ge 2$, which is set by causality 
demanding that the 
 correlation function of the magnetic field vanishes at large distances, and 
 the fact that 
 the magnetic field is divergence-free \cite{DurKahYet98}.
 In the simulations presented we chose the slope of the power spectrum 
 to be $n = 2$.  We also chose $k_c = 30$, unless specified otherwise, 
 which gives a power spectrum peaked at a relatively large value of $k$.
Biskamp and M\"uller \cite{BisMul99,MulBis00} started with a spectrum 
peaked at $k_c = 4$, which 
may account for the different slope in the late-time power spectrum which we 
observe (see Section \ref{s:specev}).
 
 Our velocity power spectrum was chosen in a similar way, but with  
 $n=0$ corresponding to white noise at large scales (there is no requirement for
 incompressibility in the early Universe).  
 The initial magnetic energy was taken equal to the kinetic energy, and 
 had the value $5\times10^{-3}$ in all runs, as the primordial field is 
 thought likely to be weak.
  
 In order to introduce a non-zero average magnetic helicity into the system it is useful 
 to represent the vector potential in terms of its projection onto an orthogonal
 basis formed by $\hat{\be}_{+}$, $\hat{\be}_{-}$ and $\hat{\bk}$.
 The two basis vectors $\hat{\be}_{+}$ and $\hat{\be}_{-}$ can be chosen
 to be the unit vectors for circular polarization, right-handed and left-handed 
 respectively. That is
 $\hat{\be}_{\pm} = \hat{\be}_{1} \pm i\hat{\be}_{2}$
 where $\hat{\be}_{1}$ and $\hat{\be}_{2}$ are unit vectors orthogonal to 
 each other and to $\bk$. They are given by 
 $\hat{\be}_{1} = \bk\times\hat{\bz}/|\bk\times(\bk\times\hat{\bz})|$ and
 $\hat{\be}_{2} = \bk\times(\bk\times\hat{\bz})/|\bk\times\hat{\bz}|$
 respectively. $\hat{\bz}$ is a reference direction.
 
 Note that since  
 \ben
   i\hat{\bk}\times\hat{\be}_{s} = sk\hat{\be}_{s}
 \een 
 where $s = \pm 1$, this corresponds to an expansion of the magnetic
 vector potential into helical modes.
 
 Using these basis vectors it is easily seen that the magnetic energy 
 spectrum is
 \ben
 E_{\rm M}(k) = 2\pi k^{2}\langle|\bB_{\bk}|^{2}\rangle
 \een
 where the amplitude of the magnetic field is given by
 \ben
   | \bB_{\bk}|^{2} = 
   (| A_{\bk}^{+}|^{2} + 
   | A_{\bk}^{-}|^{2}) |\bk|^{2}
 \een 
 and the expression for the magnetic helicity spectrum $H_{\rm M}(k)$ is
 \ben
 H_{\rm M}(k) = 4\pi k^{2}\langle\bA_{\bk}^{*}\cdot\bB_{\bk}\rangle
 \een
 where
 \ben   
   \bA_{\bk}^{*}\cdot\bB_{\bk} = 
   (| A_{\bk}^{+}|^{2} - 
   | A_{\bk}^{-}|^{2}) |\bk|.
 \een
 The function
 $H_{\rm M}(k)$ is a sensitive measure of the correlation between the vector potential
 and the magnetic field. $H_{\rm M}(k)$ may, of course, be positive in one part
 of Fourier space and negative in another part. It is, however, bounded in 
 magnitude by the inequality 
 \ben
   | H_{\rm M}(k)|\leq 2k^{-1} E_{\rm M}(k).
 \een 
 A field which saturates the above inequality is maximally helical.  
 
The amplitudes $A_{\bk}^{\pm}$ can be chosen independently, provided
$A^{*\pm}_{-\bk} = A^{\pm}_{\bk}$, which is just the condition that
the vector potential be real.
 Therefore it is possible to adjust the amplitudes $| A_{\bk}^{+}|$ and 
 $| A_{\bk}^{-}|$ freely and in so doing obtaining a magnetic field with 
 arbitrary magnetic helicity. With our 
 method we are able to put statistically random but maximally helical fields 
 in our initial conditions.  In our runs with initial helicity we take 
 $H_{\rm M} = H_{\rm max}$.
  
 
Because we evolve our dynamical fields on a discrete lattice we have to
be careful when using derivative operations in Fourier space. 
In general, the wave vector, which is an eigenvalue of the 
derivative operator,  needs to be replaced by some function
$k_{\rm eff}(k)$, which is an eigenvalue of the discrete derivative 
operator on the lattice. In our case we have, for the sixth order explicit 
centered derivative
\ben
k_{\rm eff}(k) = {\textstyle\frac{1}{30}}
\left[ \sin(3k) - 9 \sin(2k) + 45 \sin(k) \right].
\een  
In order to be consistent with the scheme used in the simulation, we use
$k_{\rm eff}(k)$ when calculating the initial condition in Fourier space.

\section{Results}
In all runs the mean kinematic viscosity $\nu$ and the resistivity $\eta$
were chosen to be equal 
with values between $\nu=\eta=5\times10^{-4}-5\times10^{-5}$.
In our simulations we typically obtain Reynolds numbers 
of the order of $100-200$. 
The Reynolds numbers in our simulations are 
evaluated using the magnetic Taylor microscale which 
we calculate here as the ratio of the rms magnetic field and the rms current 
density, $L_{\rm T} = 2\pi B_{\rm rms}/J_{\rm rms}$. The $2\pi$ factor is
here included so that $L_{\rm T}$ represents the typical wave length (and not
the inverse wavenumber) of structures in the current density.

\subsection{Spectral evolution}
\label{s:specev}
\begin{center}
\begin{figure}[p]                
\scalebox{0.4}{\includegraphics{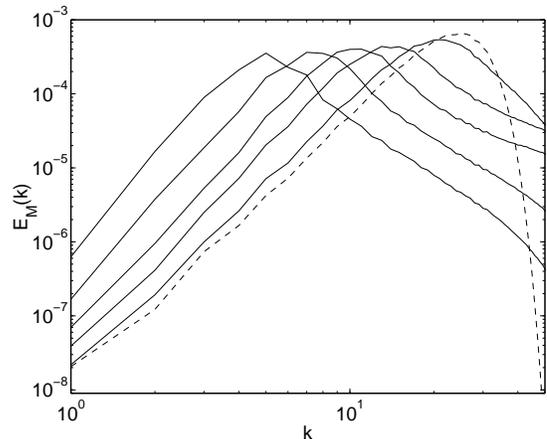}}  
\caption{Magnetic energy spectrum $E_{\rm M}(k)$ for a run with finite magnetic 
helicity. $\nu=\eta=5\times 10^{-5}$. The times shown are 
$0, 1.0, 4.6, 10.0, 21.5$ and $46.3$. The initial spectrum is indicated by 
the dashed line. 
At low wavenumbers $k$ the energy spectrum
$E_{\rm M}(k)$ increases with time.}
\label{spec.mag.hel}
\end{figure}
\end{center}

The inverse magnetic cascade for decaying MHD turbulence is best  
visualized in terms of magnetic energy spectra $E_{\rm M}(k)$ because 
information on nonlinear interaction between different scales is 
contained in $E_{\rm M}(k)$.
In Fig.~\ref{spec.mag.hel} we show a run with initial magnetic helicity.
In Fig.~\ref{spec.mag.hel} we see evidence for a dual energy transfer
both toward higher and lower wavenumbers.
The inverse cascade is characterized by the transfer of energy from small 
scale structures in the magnetic field to larger ones. 
In Fig.~\ref{spec.mag.hel} this behavior is clearly seen as 
indicated by the rise in the energy spectrum at small wavenumbers. 
Some energy is also being transported to smaller scales where
the spectrum is decaying due to diffusive effects. 
We also note that at wavenumbers above the peak $k_{p}(t)$ the spectrum 
develops a power law shape. This power law has approximately a $k^{-2.5}$
slope. 
This differs from the approximately $k^{-5/3}$ law found by M\"uller and Biskamp 
\cite{MulBis00}. We suggest that this is due to finite size effects, which
affect the spectrum if the initial scale separation between $k_{p}$ and the
smallest wavenumber in the box ($k = 1$) is insufficient, and if the flow is
strongly helical so that its spectrum is governed by inverse cascading. In
order to check this we have performed 
a run with larger initial length scale, $k_{c} = 5$. In this
case the magnetic energy spectrum develop into an approximate $k^{-5/3}$ law at
late times. However, this occurs only after the peak of the spectrum has left
the simulation box,
i.e. after finite size effects have begun to play a role. 

To check if the magnetic field evolution is self-similar one can make the 
following ansatz for the energy spectrum
\ben
E_{\rm M}(k,t) = 
\xi(t)^{-q} g_{\rm M}(k\xi)
\label{self-similar}.
\een
Here $\xi$ is the characteristic length scale of the magnetic field,
taken to be the magnetic Taylor microscale defined above, and $q$ is a parameter 
whose value is some real number.
We call $g_{\rm M}(k\xi)$ the magnetic scaling function.
\begin{center}
\begin{figure}[p]                     
\scalebox{0.4}{\includegraphics{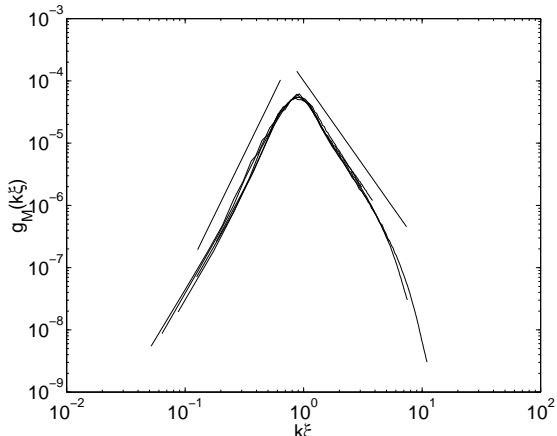}}      
\caption{The magnetic scaling function $g_{\rm M}(k\xi)$ described in the text,
equation (\ref{self-similar}), versus $k\xi$. The straight lines indicate
the power laws $\propto (k\xi)^{4.0}$ and $\propto (k\xi)^{-2.5}$ respectively.}
\label{collapse}
\end{figure}
\end{center}
In Fig.~\ref{collapse} we have plotted $\xi(t)^{q} E_{\rm M}(k,t)$ versus the 
scaled variable $k\xi(t)$. 
The value of the parameter $q$ in this run is $q=0.7$.
It is seen that for each different value of time $t$, the data collapses 
onto a single curve given by the scaling function $g_{\rm M}(k\xi)$, demonstrating the 
self-similarity of the magnetic field evolution.  

\begin{center}
\begin{figure}[p]                     
\scalebox{0.4}{\includegraphics{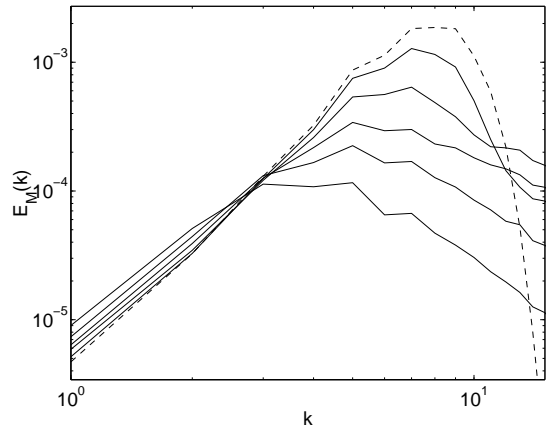}}      
\caption{Magnetic energy spectrum $E_{\rm M}(k)$ for a run with no net magnetic 
helicity. $\nu=\eta=1\times 10^{-4}$. Here $k_{c}=10$. The times shown are 
$0, 2.2, 4.6, 10.0, 21.5$ and $ 46.3$. The initial spectrum is indicated by 
the dashed line.
The peak of the energy spectrum 
$E_{\rm M}(k)$ is decreasing with increasing time.}
\label{bps_nohel}
\end{figure}
\end{center}

We also performed runs in which the magnetic helicity was zero, in the statistical 
sense. Magnetic helicity was present due to fluctuations, but was of very small 
amplitude. In these runs no significant inverse cascade was observed. 
Fig.~\ref{bps_nohel} shows the energy spectrum for such a run 
with only small magnetic helicity fluctuations present in the initial conditions. 
It is seen that only a weak inverse cascade 
is present at the lowest wavenumbers, much smaller than in the helical case. 
However, that it seem to be present at all is interesting as the effect could 
become more pronounced for higher Reynolds numbers.
It is possible that this effect is due to the magnetic helicity fluctuations 
even though they were small. 
One simulation was performed with identically zero initial magnetic helicity
fluctuations. In this case random fluctuations develop rapidly
and no differences between the two cases, were observed.

\subsection{Energy decay}
\begin{center}
\begin{figure}[p]                      
\scalebox{0.4}{\includegraphics{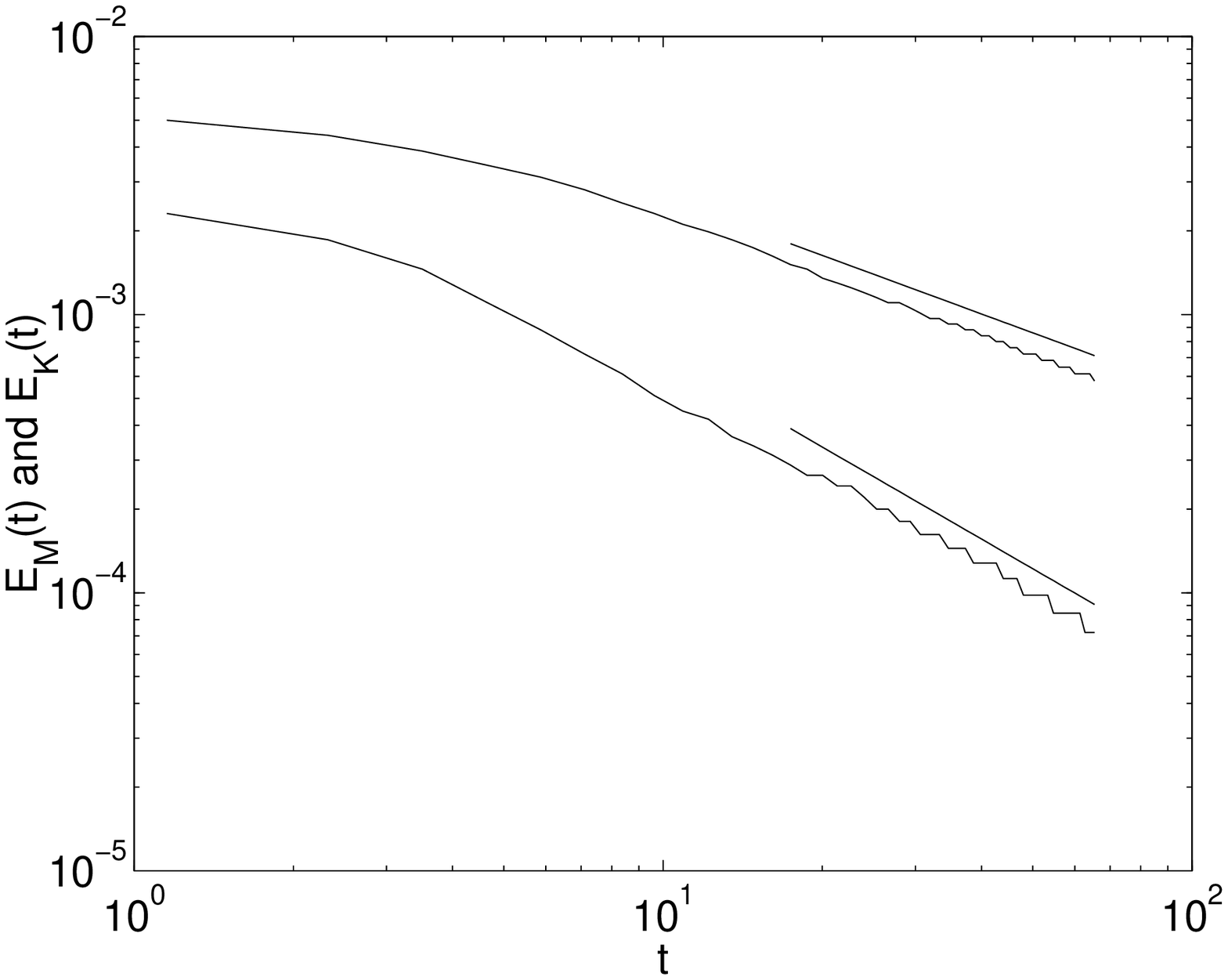}}
\caption{Time evolution of the magnetic energy $E_{\rm M}(t)$ and 
the kinetic energy $E_{\rm K}(t)$ in the case where there is 
initial magnetic helicity. $\nu=\eta=5\times 10^{-5}$.
The straight lines indicate the power laws $\propto t^{-0.7}$ and 
$\propto t^{-1.1}$ respectively.}
\label{brms_hel}
\end{figure}
\end{center}

In Fig.~\ref{brms_hel} we show the time evolution of the magnetic  
energy $E_{\rm M}(t)$ and the kinetic energy $E_{\rm K}(t)$
for a run with initial helicity and a
$k^{4}$ initial energy spectrum slope. 
It is seen that the asymptotic decay rate for $E_{\rm M}(t)$ is 
approximately $t^{-0.7}$. The Reynolds number for 
this run was around $Re\sim 200$ at late times. In another run with 
$Re\sim 100$ the decay rate was seen to be $t^{-0.8}$, so 
there seems to be a dependence of the decay rate of the magnetic field
on the Reynolds number and perhaps the resulting slope of the spectrum.

The kinetic energy also decays with a power law behavior at late times.
In the case of runs with initial helicity the kinetic energy
$E_{\rm K}(t)$ decays with a different, faster rate than $E_{\rm M}(t)$. 
The asymptotic decay rate is close to $t^{-1.1}$. 
In runs without initial helicity the decay rates of $E_{\rm M}(t)$ and
$E_{\rm K}(t)$ are approximately the same, close to $t^{-1.1}$.

In our runs with $E_{\rm K} = E_{\rm M}$ initially, the kinetic energy spectrum shows 
no evidence of an inverse cascade at any scale. 
However, when the initial velocity distribution
is zero the kinetic spectrum grows on all scales initially and in the low 
wavenumber region the energy continues to grow even after the high wavenumber 
modes start to decay.

\subsection{Coherence length evolution}
During the course of the simulations the initially small scale structures
gain in size. A convenient length scale is the magnetic Taylor microscale
$L_{\rm T}$, which was defined above. This length scale is mostly 
characteristic of the small scales but even they grow during the course of 
the simulations.

\begin{center}
\begin{figure}[p]                            
\scalebox{0.4}{\includegraphics{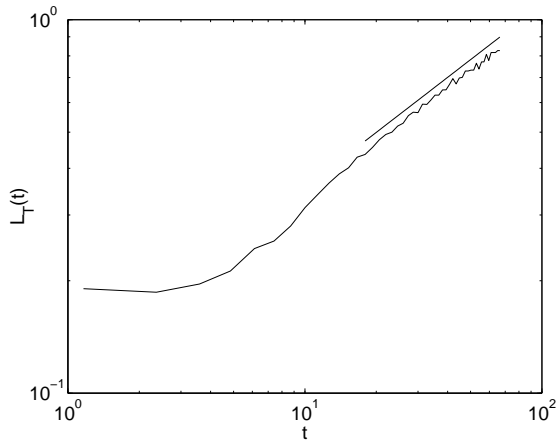}}
\caption{Time evolution of the magnetic Taylor microscale for the case
with initial magnetic helicity. $\nu=\eta=5\times 10^{-5}$.
The straight line indicates the power law $\propto t^{0.5}$.}
\label{taylor_hel}
\end{figure}
\end{center}

In Fig.~\ref{taylor_hel} we show the evolution of $L_{\rm T}$ 
for a run with initial helicity.
The asymptotic behavior of the length scale is seen to grow 
approximately as $L_{\rm T}\sim t^{0.5}$. 

In runs with non-helical initial conditions the growth of the magnetic Taylor
microscale is slower than in the case of helical initial conditions.
In this case the magnetic Taylor microscale grows approximately as 
$L_{\rm T}\sim t^{0.4}$. 

The discussion so far has mainly been concerned with the evolution of causally 
generated magnetic fields using an initial $k^{4}$ slope in the magnetic energy 
spectrum. 
Now we briefly comment on the other cases we have looked at.
For a white noise initial spectrum $E_{\rm M}(k)\sim k^{2}$, the evolution is
qualitatively and quantitatively similar to the causal case. For helical fields 
we observe an inverse cascade, while for non-helical fields a much smaller
inverse cascade is present only for the lowest modes.

\section{Discussion}
Our simulations show the decay rate of magnetic energy for compressible turbulence 
being sensitive to the initial helicity of the magnetic field configuration. 
A similar result was found in \cite{BisMul99} in the case of incompressible 
turbulence. 
The fact that magnetic helicity is conserved (except for resistive changes),
and the magnetic energy decays slower for helical fields, is connected with
the observed inverse cascade in which magnetic energy is transported toward
larger scales because of nonlinear dynamics.

The decay of kinetic energy does not seem to depend on the initial helicity and its
decay rate $E_{K}(t)\sim t^{-1.1}$ is consistent with the earlier work of 
\cite{Mac+98,BisMul99}. Note that in the helical case we observe the 
kinetic energy decaying more rapidly than the magnetic one; 
this behavior was also found in \cite{BisMul99}.

While these results are not directly applicable to the 
evolution of primordial magnetic fields in the early universe, they do suggest that 
nonlinear magnetohydrodynamical effects may play an important role in this case.

In any case, it is interesting to compare our results with the work of other authors 
interested in the decay properties of cosmological magnetic fields  
\cite{Ole97,Son98,Shi98,FieCar98}.  Ideal MHD has a scale invariance which 
leads to the scaling law \cite{Ole97,Shi98}
\ben
E_M(t,k) = k^{-1-2h}\psi(k^{1-h}t),
\een
where $\psi$ is an unknown function, related to $g_{\rm M}$. 
Assuming it is peaked somewhere, and $h<0$,
the characteristic scale of the field goes as $L(t) \sim t^{1/(1-h)}$.  It is 
also often assumed that $\psi(0)$ exists and is non-zero: thus $h$ is 
determined by the initial power spectrum.  Hence for a magnetic power spectrum
of index $n$, $h=-(n+3)/2$ and 
\ben
\label{e:lenscalaw}
L(t) \sim t^{2/(n+5)}.
\een
This law can also 
be recovered by assuming that the characteristic time scale for the decay of 
turbulence on a scale $l$ is the eddy turnover time $\tau_l = l/v_l$, where 
$v_l \sim l^{-(n+3)}$ is the velocity averaged on a scale $l$ \cite{Son98}. 
If the characteristic scale of the field is that scale which is just decaying, 
then $\tau_L \sim t$, and we again find Eq. (\ref{e:lenscalaw}).  
One should note that these arguments ignore helicity conservation.  

We recall that our non-helical runs had $n=2$ for the magnetic power 
spectrum and $n=0$ for the velocity power spectrum. The observed growth law 
for the magnetic Taylor microscale, $t^{0.4}$, is not consistent with 
the predicted power law for $n=2$, although it does square with the growth 
law for $n=0$, and it is possible that the growth in the magnetic field 
length scale is being controlled by the velocity field. Simulations at higher
Reynolds numbers seem required to resolve this issue.

One would expect on integrating the helicity power spectrum that $H_M \sim L_I E_M$, 
where $L_I$ is the integral scale. We would expect that $L_I \sim L_T$, and hence, 
if magnetic helicity is conserved, 
\ben
\label{e:helELreln}
E_M \sim L_T^{-1}.  
\een
However, magnetic helicity is not conserved exactly: we observe a decrease in
$H_{\rm M}$ by a factor of about 2 in a run with viscosity 
$\nu=5 \times 10^{-5}$.
Indeed, with $L_T \sim t^{0.5}$ we find a somewhat steeper relation:
$E_M \sim t^{-0.7}\sim L_T^{-1.4}$.

Finally, it is interesting to note that Son's numerical simulations of 
decaying turbulence \cite{Son98}, performed in the Eddy-Damped Quasi-Normal Markovian 
(EDQNM) approximation, show some evidence of a power law developing at high $k$,
the slope being close to $k^{-2.5}$,
although there was no net helicity present, and no inverse cascade.
Furthermore, Field and Carroll \cite{FieCar98}, again using the EDQNM
approximation, found that there were self-similar solutions with 
$E_M \sim t^{-2/3} \sim L_T^{-1}$.

\section{Conclusions}
We have shown that for an isothermal and compressible magnetized turbulent fluid, 
when undergoing a process of free decay, a substantial inverse cascade 
is present for helical magnetic field configurations,
which transfer energy from smaller scale magnetic fluctuation to larger scale ones.
For non-helical magnetic fields only a weak inverse cascade was
observed on the largest scales. 

The energy spectrum of the magnetic field shows evidence for a self-similar evolution
with a development of a power law of roughly $k^{-2.5}$ beyond the peak.
Decay laws for both the kinematic and magnetic energy were found. The kinetic energy 
decay was approximately $t^{-1.1}$ for both helical and non-helical magnetic fields. 
The decay of the magnetic field energy was found to be strongly dependent on the the 
initial helicity, decaying roughly as $t^{-0.7}$ and $t^{-1.1}$ for helical and 
non-helical initial conditions respectively. For the helical case, the magnetic energy
decay rate showed a dependence on the Reynolds number, with a slower decay rate for 
larger Reynolds numbers.   

We also observed power law behavior in the characteristic length scale of the 
magnetic field, defined as the Taylor microscale $L_{T}$. In the helical case
$L_{T} \sim t^{0.5}$, whereas for non-helical fields the growth was somewhat slower, 
$L_{T} \sim t^{0.4}$, and we ascribe the faster growth rate to the presence of the
inverse cascade in the helical case.

\acknowledgements
This work was conducted on the Cray T3E and SGI Origin platforms using COSMOS 
Consortium facilities, funded by HEFCE, PPARC and SGI. We also acknowledge
computing support from the Sussex High Performance Computing Initiative.

    

\end{document}